\documentclass{aa}
\usepackage{graphicx}
\usepackage{txfonts}
\begin{document}
   \title{Asteroseismology of the visual binary 70~Ophiuchi\thanks{Based
   on observations collected at the 3.6-m telescope at La Silla Observatory (ESO, Chile: program 073.D-0590)}}

   \author{F. Carrier
          %\inst{1}
          \and
          P. Eggenberger
	  %\inst{1}
          }

   \offprints{F. Carrier\\
      \email{fabien.carrier@obs.unige.ch}}

   \institute{Observatoire de Gen\`eve, 51 chemin des Maillettes,
             CH-1290 Sauverny, Switzerland 
             }

   \date{Received ; accepted }

% \abstract{}{}{}{}{} 
% 5 {} token are mandatory
 
  \abstract
   {Convection in stars excites resonant acoustic waves. 
The frequencies of these oscillations depend on the sound speed
inside the star, which in turn depends on density, temperature, gas motion,
and other properties of the stellar interior. Therefore, analysis
of the oscillations provides an unrivaled method to probe the
internal structure of a star.}
   {Solar-like 
oscillations in the primary of the visual 
binary 70~Ophiuchi are investigated.}
   {70~Ophiuchi~A was observed with the \textsc{Harps} spectrograph 
   mounted on the 3.6-m telescope at the ESO La Silla Observatory (Chile) during 6 nights
   in July 2004 allowing us to collect 1758 radial velocity measurements with a standard 
   deviation of about 1.39\,m\,s$^{-1}$.}
   {The power spectrum of the
high precision velocity time series clearly presents several identifiable peaks between 3 and
6\,mHz showing regularity with a large spacing of $\Delta\nu$\,=\,161.7\,$\pm$\,0.3\,$\mu$Hz. 
Fourteen individual modes were
identified with amplitudes in the range 11 to 14\,cm\,s$^{-1}$.}
   {}


   \maketitle
%
%________________________________________________________________

\section{Introduction}

Analysis of the oscillation spectrum provides an unrivaled method to probe the stellar
internal structure, since the frequencies of these oscillations depend on the sound speed inside the
star, which in turn depends on density, temperature, gas motion, and other properties of the stellar
interior. The prime example of such a probe is the Sun: the five-minute oscillations
have provided a wealth of information
about the solar interior. The Sun oscillates simultaneously in many modes with
periods of about 5 minutes and Doppler amplitudes of about 23\,cm\,s$^{-1}$ for
the strongest modes. Observation of thousands of oscillation frequencies and
comparing them with theoretical calculations has led to significant revisions
of solar models.

These results stimulated various attempts to detect a similar
signal on other solar-like stars and the stabilized spectrographs have
achieved the
accuracy needed
for solar-like oscillation detection by means of radial velocity measurements (Carrier et al. 
\cite{cbe03}). However, a major asteroseismological difficulty is the confrontation of observations
and theoretical models (Eggenberger et al. \cite{ecte04}, \cite{ecb05}). The observational 
measurements available for an isolated star - such as the effective temperature,
the metallicity, the luminosity, and eventually the interferometric radius, combined 
with oscillation frequencies -
provide strong constraints to the global parameters of the star but
are often not sufficient to 
unambiguously determine an adequate model and to really test the physics of the models.
The additional constraints imposed by the binary nature, namely the same age and initial
chemical composition, are extremely valuable for accurately determining the
properties of a binary system. Moreover, in the case of binaries, 
the masses of both components are accurately known by combining visual and spectroscopic orbits.
A primary target for the search for p-mode oscillations in such a system is the bright K0 dwarf
\object{70~Ophiuchi A} (\object{HD~165341A}),
in addition to the $\alpha$~Cen system (Bouchy \& Carrier \cite{bc02}, Carrier \& Bourban \cite{cb03}, 
Bedding et al. \cite{bkbe04} and Kjeldsen et al. \cite{kbbe05})
and Procyon (Eggenberger et al. \cite{ecbb04}, Martic et al. \cite{mlak04}).

In this paper, we report Doppler observations of 70~Ophiuchi~A made with the \textsc{Harps} spectrograph
resulting in the detection of p-mode oscillations. 
The observations and data
reduction are presented in Sect.~\ref{odr}, 
the acoustic spectrum analysis and the mode identification in Sect.~\ref{psa},
and the conclusion is given in Sect.~\ref{conclu}.
%__________________________________________________________________

\section{Observations and data reduction}
\label{odr}
With a $V$-magnitude of 4.1, 70~Oph A
is the main component of a spectroscopic visual binary composed of a K0 and a K5 dwarf with an 
orbital period of 88.38\,yr (Pourbaix \cite{p00}).
It was observed over 6 nights in July 2004 with the \textsc{Harps} spectrograph (Pepe et al. \cite{pmr02}) mounted on the 3.6-m
telescope at La Silla
Observatory (ESO, Chile). We took sequences of 40-60\,s exposures, depending on the airmass and the extinction, 
with a dead time of 30\,s in-between.
In total, 1758 spectra were collected with a typical signal-to-noise ratio (S/N) in the range of 120-220 at 530\,nm.
During the stellar exposures, the spectrum of a thorium lamp carried by a second fiber was simultaneously recorded
in order to monitor the spectrograph's stability.
The spectra obtained were extracted on-line. The radial velocities
were computed by weighted cross-correlation with a numerical mask constructed from a K5 dwarf spectrum. They were also
determined by the optimum-weight procedure (Connes \cite{c85}, Carrier et al. \cite{cbk01}) but without significant gain.

   \begin{figure}
   \resizebox{\hsize}{!}{\includegraphics{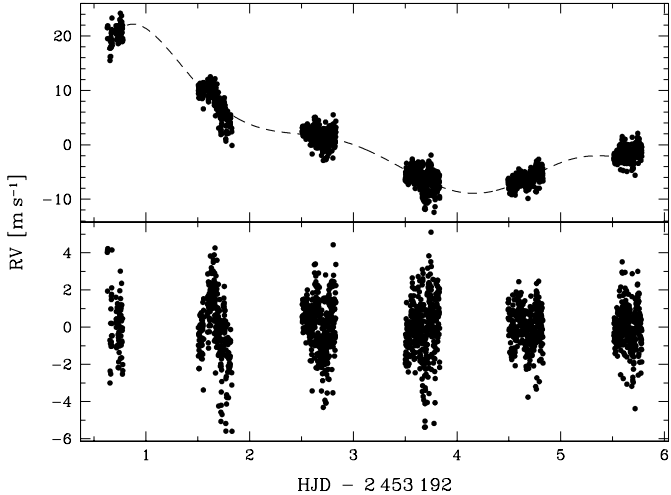}}
   \caption{Radial velocity measurements of 70~Oph~A. {\bf Top: } A constant value of -5.71\,km\,s$^{-1}$ was
   removed from the original data. {\bf Bottom: } An 8-order polynomial fit was
   subtracted (dashed line on the top window).}
              \label{figrv}%
    \end{figure}

The modulation of the Doppler measurements is due to the binarity of the star (see Fig.~\ref{figrv}). Thus, an
8-order polynomial fit was subtracted from the raw data to best eliminate the Keplerian movement and variations caused by
temperature fluctuations of the instrument during a few days due to a power failure that happened just
before our observation run. This high-pass
filtering does not affect the results of the p-modes analysis: indeed, a 2 or an 8-order polynomial fit lead to
the same power spectrum above 2\,mHz.
The rms scatter of this "new" time series is 1.39\,m\,s$^{-1}$ (see Table~\ref{tabrv}), the solar-like
oscillations representing a non-neglecting part of this number.
\begin{table}
\caption{Distribution and dispersion of Doppler measurements.}
\begin{center}
\begin{tabular}{clll}
\hline
\hline
Date & No spectra & No hours & $\sigma$ (m\,s$^{-1}$) \\ \hline
2004/07/05 & 81  & 3.58  & 1.54 \\
2004/07/06 & 290 & 7.78  & 1.76 \\
2004/07/07 & 350 & 7.80  & 1.39 \\
2004/07/08 & 390 & 7.95  & 1.52 \\
2004/07/09 & 331 & 8.07  & 1.00 \\
2004/07/10 & 316 & 6.45  & 1.11 \\ \hline
\label{tabrv}
\end{tabular}
\end{center}
\end{table}

\section{Power spectrum analysis}
\label{psa}
\begin{figure}
   \resizebox{\hsize}{!}{\includegraphics{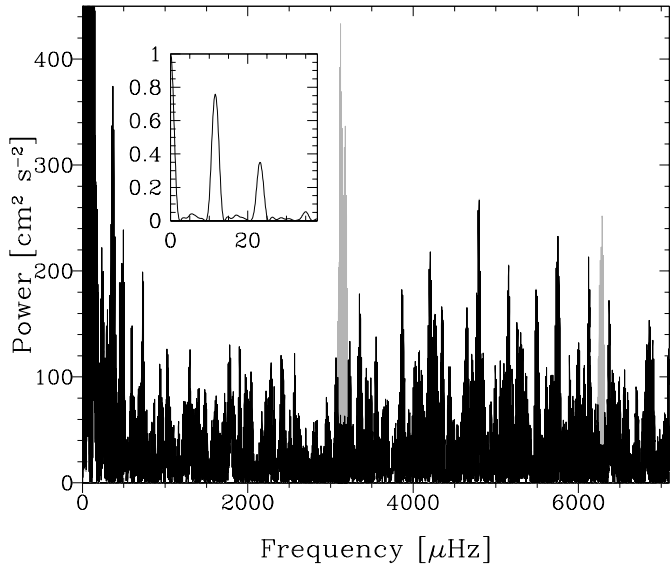}}
   \caption{Power spectrum of 70~Oph~A. The high peak in gray (and twice this value) should be due to a technical problem (probably due to a hard point
   of the telescope). The Nyquist frequency has a value of 7.14\,mHz. The window function is shown
   in the inset with daily aliases at 11.57 and 23.15\,$\mu$Hz.}
              \label{figtf}%
    \end{figure}

In order to compute the power spectrum of the velocity time series, we use
the Lomb-Scargle modified algorithm (Lomb \cite{lomb}; Scargle \cite{scargle}). Its time scale gives a formal resolution
of 2.2\,$\mu$Hz. The resulting periodogram, shown in Fig.~\ref{figtf},
exhibits a series of peaks between 3 and 6\,mHz, exactly where
the solar-like oscillations are expected for this star. 
Typically for such a power spectrum, the noise has two components:
\begin{itemize}
\item At high frequencies it is flat, indicative
of the Poisson statistics of photon noise. The mean white noise level $\sigma_{\mathrm{pow}}$ calculated between 1.5 and 2.5~mHz 
is 0.0026\,m$^2$\,s$^{-2}$, namely
$\sigma_{\mathrm{amp}} = \sqrt{\sigma_{\mathrm{pow}} * \pi / 4}$\,=\,4.5\,cm\,s$^{-1}$ in amplitude (Kjeldsen \& Bedding \cite{kb}). With 1758 measurements, this high frequency noise 
corresponds to $\sigma_{RV}\,=\,\sqrt{N \sigma_{\mathrm{pow}} /4 }\,=\,1.07$~m\,s$^{-1}$. This radial velocity uncertainty is
larger than the uncertainty due to photon noise, which is estimated at 0.25-0.4\,m\,s$^{-1}$. This discrepancy could
be explained by the bad seeing and the technical problems occurring during the observation run.
However, part of the noise should have a stellar origin (like granulation).
\item Towards the lowest frequencies, the power should scale inversely with frequency squared, as expected for instrumental instabilities. 
Moreover, this star shows low frequencies, not completely removed by the polynomial fit, due to the Keplerian movement.
\end{itemize}

We note that a high peak near 3.1\,mHz and a smaller one nearly twice this value
are widely decentered with regard to the others (see Fig.~\ref{figtf}). The ESO staff pointed out that a large noise peak could occur with 
a period of about 6 minutes (near 3\,mHz and multiple frequencies at 6 and 9\,mHz). This noise is not completely understood at present; it could be due
to the guiding, although this seems unlikely given that the guiding correction was applied every 5 seconds,
or to the telescope. Afterwards, we thus limit our p-modes research between 3.2 and 6\,mHz.
    
\subsection{Search for a comb-like pattern}
\begin{figure}
   \resizebox{\hsize}{!}{\includegraphics{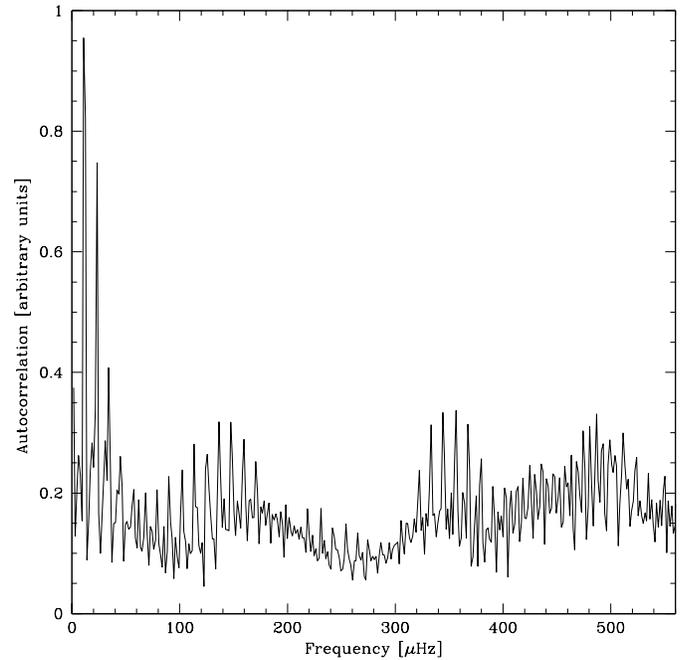}}
   \caption{Autocorrelation of the power spectrum of 70~Oph~A with a threshold of 80\,cm$^2$\,s$^{-2}$.}
              \label{figauto}%
    \end{figure}

In solar-like stars, p-mode oscillations of low-degree are expected to produce a
characteristic comb-like structure in the power spectrum with mode
frequencies
$\nu_{n,\ell}$ reasonably well approximated by the asymptotic
relation (Tassoul \cite{tassoul80}):
\begin{eqnarray}
\label{eq1}
\nu_{n,\ell} & \approx &
\Delta\nu(n+\frac{\ell}{2}+\epsilon)-\ell(\ell+1) D_{0}\;.
\end{eqnarray}
Here, $D_0$, is sensitive to the sound speed near the core and is equal to
$\frac{1}{6} \delta\nu_{02}$ when the 
asymptotic relation holds exactly; and $\epsilon$ is sensitive to
the surface layers. 
The quantum numbers $n$ and $\ell$ correspond to the radial
order and the angular degree of the modes, and $\Delta\nu$ and
$\delta\nu_{02}$
to the large and small separations.
To search for periodicity in the power spectrum, an autocorrelation with a threshold of 80\,cm$^2$\,s$^{-2}$
(all values below this limit are fixed to zero)
is calculated and presented in Fig.~\ref{figauto}.
Each peak of the autocorrelation corresponds to a structure present in the power spectrum.    
The three strong peaks at low frequency at about 11.5, 23, and 35\,$\mu$Hz correspond to the daily aliases.
The next dominant peaks appear between 120 and 170\,$\mu$Hz, and one of them corresponds to the large spacing. 
Note that the external parameters of 70~Oph~A,
scaling from the solar case (Kjeldsen \& Bedding \cite{kb}), give a large frequency spacing between 150 and 170\,$\mu$Hz.
Indeed, by refining the orbit of the system and measuring the individual magnitudes and
effective temperatures (for both components), we obtain the following parameters for 70~Oph~A:
a mass near 0.87\,M$_{\odot}$, a luminosity of 0.52\,L$_{\odot}$, an effective temperature of
5300\,K, and a solar metallicity.
Details about these parameters are postponed to the theoretical paper (Eggenberger et al. in preparation).
Other important peaks in the autocorrelation are situated between 320 and 380\,$\mu$Hz,
corresponding to twice the large separation. The large spacing will then correspond,
knowing the domain of one time and twice the large spacing, to half of 333.8\,$\pm$\,11.57\,$\mu$Hz,
namely the two strong peaks at about 161.5 and 172\,$\mu$Hz and the small one at 167\,$\mu$Hz.
To consolidate our solutions we selected all those peaks between 3.2 and 6\,mHz with an amplitude greater than 10\,cm\,s$^{-1}$ and
tried to build echelle diagrams with different large separations. Solutions showing well-aligned frequencies
are found for large spacings of about 161.5, 167, and 172\,$\mu$Hz.

\subsection{Mode identification}
The method used to extract mode frequencies is described well in Carrier et al (\cite{ceb05},\cite{cedw05}).
The frequencies are extracted using an iterative algorithm that identifies the highest peak between 3.2 and 6\,mHz
and subtracts it from the time series. Note that because of the stochastic nature of solar--like oscillations, a timestring
of radial velocities cannot be expected to be perfectly reproduced by a sum of sinusoidal terms. Therefore, using
an iterative clean algorithm to extract the frequencies can add additional peaks with small amplitudes due to the finite
lifetimes of the modes that we do not know. In the case of \object{70~Oph~A}, 
we iterated the process until all peaks with an amplitude higher than 2.5\,$\sigma$ in the amplitude
spectrum were removed. Then, $\sigma$ represents the noise in the amplitude spectrum and has a value of 
4.5\,cm\,s$^{-1}$ (see Sect.~\ref{psa}).
This threshold ensures that the selected peaks have
only a small chance of being due to noise.

Using the extracted frequencies, we drew echelle diagrams with different large spacings. The best echelle diagrams
are found for the large spacings of 161.7 and 172.2\,$\mu$Hz. The solution with a large spacing of 167\,$\mu$Hz, found
in the previous section, is ruled out by the echelle diagram; indeed, the distance between the corresponding
$\ell$\,=\,0 (or 2) and $\ell$\,=\,1 modes is not compatible with such a large separation.
The results of the automatic extraction are presented in Table~\ref{tab:dentif} for the two different possible
large separations with the corresponding identification. Because of the daily alias of 11.57\,$\mu$Hz introduced by the mono-site observations, 
we cannot know a priori whether the frequency
selected by the algorithm is the right one or an alias. We thus considered
that the frequencies could be shifted by $\pm 11.57$ $\mu$Hz.

\begin{table}
\caption[]{Identification of extracted frequencies. The signs + or -- mean that the peaks have to be shifted
to a higher or lower value, respectively, by 11.57\,$\mu$Hz.}
\begin{center}
\begin{tabular}{rccc}
\hline
\hline
\multicolumn{1}{c}{Frequency} & Mode ID ($\Delta \nu_1$) & Mode ID ($\Delta \nu_2$) & S/N \\
\multicolumn{1}{c}{$[\mu$Hz$]$} &   161.7\,$\mu$Hz        &172.2\,$\mu$Hz  &   \\
\hline
% 3055.7   & $\ell=0$           & noise      & 2.6 \\
% 3089.2   & noise              & $\ell=1$   & 2.6 \\
% 3121.2   & + $\ell=1$         & noise      & 4.6 \\  
% 3141.0   & -- $\ell=1$        & noise      & 4.2 \\
% 3168.8   & noise              & + $\ell=0$ & 4.0 \\
 3246.2   & noise              & + $\ell=1$ & 3.2 \\
 3350.9   & noise              & $\ell=0$   & 2.8 \\
 3378.1   &  $\ell=0$          & noise      & 2.6 \\
 3552.1   & -- $\ell=0$        & noise      & 2.6 \\
 3862.1   & $\ell=0$           & $\ell=0$   & 3.0 \\
 4194.4   & --	$\ell=0$       & + $\ell=0$ & 3.2 \\
 4258.8   &$\ell=1$            & noise      & 2.9 \\
 4343.7   &$\ell=0$	       & noise      & 2.8 \\
 4650.4   & + $\ell=2$	       & -- $\ell=1$& 2.8 \\
 4719.6   & noise              & $\ell=0$   & 2.5 \\
 4797.9   & noise              & + $\ell=1$ & 3.6 \\
 4993.2   & $\ell=0$	       & -- $\ell=1$& 2.5 \\
 5054.4   & + $\ell=1$         & + $\ell=0$ & 2.8 \\
 5153.7   &	$\ell=0$       & $\ell=1$   & 3.1 \\
 5255.2   & noise              & -- $\ell=0$& 2.9 \\
 5328.5   & -- $\ell=0$        & $\ell=1$   & 2.6 \\
 5487.3   & -- $\ell=2$        & + $\ell=1$ & 2.7 \\
 5717.8   &$\ell=1$            & noise      & 2.7 \\
 5763.3   & noise              & $\ell=0$   & 3.4 \\
 5890.9   & -- $\ell=1$           & noise      & 2.6 \\
 5990.0   & noise              & noise      & 2.7 \\ 
\hline
& 7 noise peaks & 7 noise peaks &\\
\hline
\end{tabular}
\end{center}
\label{tab:dentif}
\end{table}

To investigate how many peaks are expected to be due to noise, we conducted simulations in which we analyzed 
noise spectra
containing no signal. 
For this purpose, a velocity time series was built, using the observational time sampling
and radial velocities randomly drawn by assuming a Gaussian noise (Monte--Carlo
simulations).
The amplitude spectrum of this series was then calculated and peaks with
amplitude greater than 2.5,
3, and 3.5\,$\sigma$ were counted; note that a peak and its aliases are only counted once. 
The whole procedure was repeated 1000 times to ensure the
stability of the results. In this way, we find that the number of peaks due to noise with an amplitude
larger than 2.5\,$\sigma$ is 13.0\,$\pm$\,4.6 in the range 3.2--6\,mHz, 
for 3\,$\sigma$ it is 1.9\,$\pm$\,1.4, and for 3.5\,$\sigma$ the number of peaks due to noise varies between 0 and 3 
with a mean value of 0.2 and a standard deviation of 0.4. These results are in good agreement with the number of
identified noise peaks (see Table~\ref{tab:dentif}).

The star 70~Oph~A is very similar to $\alpha$~Cen~B (see Carrier \& Bourban \cite{cb03}), the same spectral type and similar large
spacing, which has a mean small spacing of 10\,$\mu$Hz in the frequency range 3-4.5\,mHz.
Inspecting the results of the mode identification, we note that the value of the small spacing coming from the identification
with the large separation of 172.2\,$\mu$Hz is significantly different from 10\,$\mu$Hz. No matter
what the identification may be ($\ell$\,=\,0, 1 or 2 for
both lines in the echelle diagram), the small separation will be lower than 6.5\,$\mu$Hz in the frequency
range 3--4.5\,mHz. This value of the small spacing, which is lower than expected, suggests that this identification
is less reliable than the one with a large separation of 161.7\,$\mu$Hz. Although the solution $\Delta\nu\,=\,172.2$\,$\mu$Hz
cannot be definitely ruled out, we afterwards consider only the one
with a large spacing of 161.7\,$\mu$Hz.

The echelle diagram based on the extracted frequencies shows only two lines.
It is not possible to disentangle $\ell$\,=\,2 from $\ell$\,=\,0 modes with our time series,
mainly due to the poor resolution (2.2\,$\mu$Hz): daily aliases of $\ell$\,=\,0 modes merge with $\ell$\,=\,2 modes,
and vice versa. This suggests that the small spacing
is close to 10\,$\mu$Hz. As a result, some extracted $\ell$\,=\,2 frequencies
could be shifted by 11.57\,$\mu$Hz and be identified as radial modes and inversely.

   \begin{figure}
   \resizebox{\hsize}{!}{\includegraphics{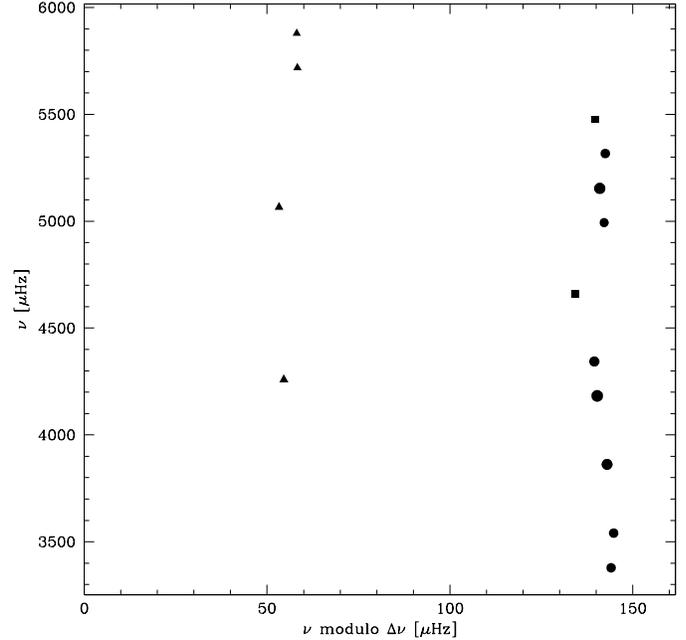}}
   \caption{Echelle diagram of identified modes with a large separation of 161.7\,$\mu$Hz. The
modes $\ell$=2 ($\blacksquare$), $\ell$=0 ({\Large $\bullet$}),
and $\ell$=1 ($\blacktriangle$) 
are represented with a size proportional to
their amplitude.}
              \label{fig:ed}%
    \end{figure}
Due to the ambiguous identification of $\ell$\,=\,0,\,2 modes, another identification can be found
replacing all $\ell$\,=\,0,2
by $\ell$\,=\,1 modes and all $\ell$\,=\,1 by $\ell$\,=\,2 modes. In this case, the peak at 4662.0\,$\mu$Hz would 
be due to noise.

\begin{figure}
   \resizebox{\hsize}{!}{\includegraphics{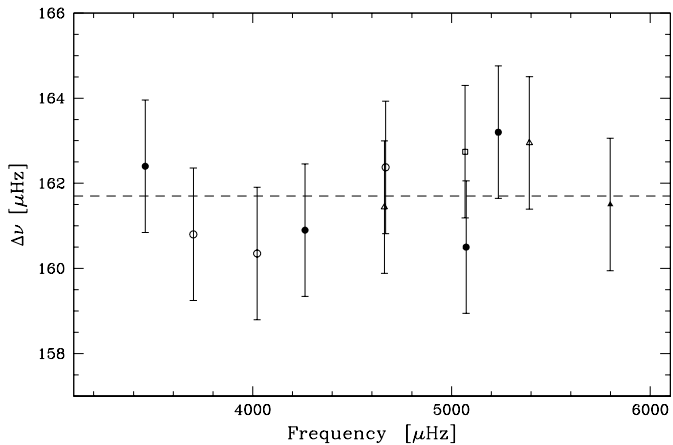}}
   \caption{Large spacing $\Delta\nu$ versus frequency for p-modes of degree $\ell$=0 ({\Large $\bullet$})
and $\ell$=1 ($\blacktriangle$). Open symbols correspond to large spacing averages taken between non-successive
modes. All individual errors are fixed to $\sqrt{2}\,\times\,1.1\,\mu$Hz (half resolution). The dashed-line corresponds to the mean large spacing of 161.7\,$\mu$Hz.}
              \label{fig:gde}%
    \end{figure}

The echelle diagram with the fourteen identified modes is shown in Fig.~\ref{fig:ed}. 
The frequencies of the modes are given in Table~\ref{tab:freq}. As the curvature of the p-mode alignment
could be important above 5.5\,mHz, both identified high-frequency
$\ell$\,=\,1 modes can take two different values shifted by 11.57\,$\mu$Hz. This should be taken into account
when comparing individual frequencies with theoretical models.
In the same way,
the peak at 5475.7\,$\mu$Hz was identified as a $\ell\,=\,2$ mode,
taking a probable curvature at high frequency into account, but could just as easily be a radial mode. 
The large separations are given in Fig.~\ref{fig:gde} and have
a mean value of 161.7\,$\pm$\,0.3\,$\mu$Hz.

\begin{table}
\caption{Oscillation frequencies (in $\mu$Hz) for the large spacing of 161.7\,$\mu$Hz. The frequency resolution
of the time series is 2.2\,$\mu$Hz, and n' is the radial order defined within a constant value. Another possible identification can be done by
inverting all
$\ell$\,=\,0,2 and $\ell$\,=\,1 modes, except for the peak at 4662.0\,$\mu$Hz, which would be due to noise in this
case.}
\begin{center}
\begin{tabular}{lccc}
\hline
\hline
n' & $\ell$ = 0 & $\ell$ = 1 & $\ell$ = 2 \\
\hline
% -9 & 3055.7* & 3131.1* &			\\
% -8 &        & 	 &			\\
 -7 & 3378.1 &        & 		     \\
 -6 & 3540.5 &        &    \\
 -5 &	     &        &        \\
 -4 & 3862.1 &        &  \\
 -3 &	     &        &        \\
 -2 & 4182.8 & 4258.8 &  \\
 -1 & 4343.7 &        & 	   \\
 0 &	    &	     &    \\
 1 &	    &	     & 4662.0  \\
 2 &	    &	     &  		   \\
 3 & 4993.2 & 5066.0 &  		   \\
 4 & 5153.7 &	     &  		   \\
 5 & 5316.9 &	     &  		   \\
 6 &	    &	     & 5475.7			   \\
 7 &	    & 5717.8  & 		 \\
 8 &	    & 5879.3  & 		 \\
\hline
\end{tabular}\\
\end{center}
\label{tab:freq}
\end{table}

\subsection{Oscillation amplitudes}
Concerning the amplitudes of the modes, theoretical computations predict oscillation 
amplitudes near 13\,cm\,s$^{-1}$ for a star 
like \object{70~Oph~A}, with mode lifetimes on the order of a few days
(Houdek et al. \cite{ho99}). The amplitudes of the highest peaks are
in the range 11--14\,cm\,s$^{-1}$, in agreement with the theoretical value and with the observations
of a similar star, $\alpha$~Cen B (Carrier \& Bourban \cite{cb03}).
\section{Conclusion}
\label{conclu}
Our observations of 70~Oph~A yield a clear detection of p-mode oscillations. Several identifiable modes appear
in the power spectrum between 3 and 6\,mHz with an average large spacing of 161.7\,$\mu$Hz and a maximal amplitude
of 14.5\,cm\,s$^{-1}$. The mono-site observations, coupled to the low resolution of the time series and to the faint
signal-to-noise, do not allow us to unambiguously disentangle $\ell$\,=\,2 from $\ell$\,=\,0 modes. The small
spacing seems to be consistent with theoretical predictions for such a star with a value near
10\,$\mu$Hz. However, the lack of couples of $\ell$\,=\,0 and 2 modes for
the same radial order does not give a single identification solution: $\ell$\,=\,0,2 can be replaced by $\ell$\,=\,1 modes
and $\ell$\,=\,1 by $\ell$\,=\,2 modes. 
The study of the 70~Oph system, with asteroseismic and non-asteroseismic constraints, has been postponed to a second paper.
\begin{acknowledgements}
      Part of this work was supported by the Swiss National Science Foundation.
\end{acknowledgements}

\end{document}